\documentclass[conference]{IEEEtran}
\ifCLASSINFOpdf
\else
\fi
\usepackage{hyperref}
\usepackage{amssymb}
\usepackage{multirow}
\usepackage{graphicx}
\usepackage{amsmath}
\usepackage[table]{xcolor}
\definecolor{lightgray}{gray}{0.9}
\usepackage{filecontents}
\usepackage{subfig}
\usepackage{array}
\usepackage{amsmath}
\usepackage{pbox}
\usepackage[linesnumbered,ruled]{algorithm2e}
\usepackage{float}
\usepackage{algorithmic}

\long\def\comment#1{}

\begin{document}
\graphicspath{ {images/} }

\title{Automated U.S Diplomatic Cables Security Classification: Topic Model Pruning vs. Classification Based on Clusters}

\author{\IEEEauthorblockN{Khudran Alzhrani\IEEEauthorrefmark{1},
Ethan M. Rudd\IEEEauthorrefmark{2}, C. Edward Chow\IEEEauthorrefmark{1} and
Terrance E. Boult\IEEEauthorrefmark{2}}
\IEEEauthorblockA{
  University of Colorado at Colorado Springs\\
  \IEEEauthorrefmark{1}\IEEEauthorrefmark{2}Department of Computer Science\\
  \IEEEauthorrefmark{2}Vision and Security Technology (VAST) Lab\\
  Email: \IEEEauthorrefmark{1} \{kalzhran,cchow\}@uccs.edu \IEEEauthorrefmark{2} \{erudd,tboult\}@vast.uccs.edu
}
Address: 1420 Austin Bluff’s Parkway, Department of Computer Science, Colorado Springs, CO, 80907\\
Phone: (719) 255-3544 Fax: (719) 255-3369
}

\maketitle

\begin{abstract}
The U.S Government has been the target for cyber-attacks from all over the world. 
Just recently, former President Obama accused the Russian government of the leaking emails to Wikileaks and declared that the U.S. might be forced to respond.
While Russia denied involvement, it is clear that the U.S. has to take some defensive measures to protect its data infrastructure. 
Insider threats have been the cause of other sensitive information leaks too, including the infamous Edward Snowden incident. 
Most of the recent leaks were in the form of text. 
Due to the nature of text data, security classifications are assigned manually. 
In an adversarial environment, insiders can leak texts through E-mail, printers, or any untrusted channels. 
The optimal defense is to automatically detect the unstructured text security class and enforce the appropriate protection mechanism without degrading services or daily tasks.
Unfortunately, existing Data Leak Prevention (DLP) systems are not well suited for detecting unstructured texts.
In this paper, we compare two recent approaches in the literature for text security classification, evaluating them on actual sensitive text data from the WikiLeaks dataset.
\end{abstract}

\IEEEpeerreviewmaketitle

\section{Introduction and Motivation}

The sensitivity of information can be assessed based on the impact that might result from its leakage.
From an international relations perspective, sensitive information leaks could damage relationships between the U.S. and its allies, as resulted from the 2011 leakage of diplomatic cables by Wikileaks and the 2013 Edward Snowden Leaks.
From a political perspective, sensitive data leaks can create scandals, sway elections, and end careers, as we saw with the 2016 DNC Email Leaks and the 2016 Panama Papers.
While these scandals have certainly garnered widespread media attention, and have been hotly debated in the mainstream media, with some calling for the incarceration of Julian Assange and Edward Snowden, and others treating these individuals as sympathetic figures. 
Regardless of one's own political opinions on these respective incidents, however, the magnitude at which such leaks affect change and the degree of press coverage that they achieve -- with little barrier to entry -- means that they pose an enormous opportunity for bad actors interested in negatively impacting the national security, the finances of institutions, and human lives.

\begin{figure}[!t]
\centering
{\includegraphics[scale=.085]{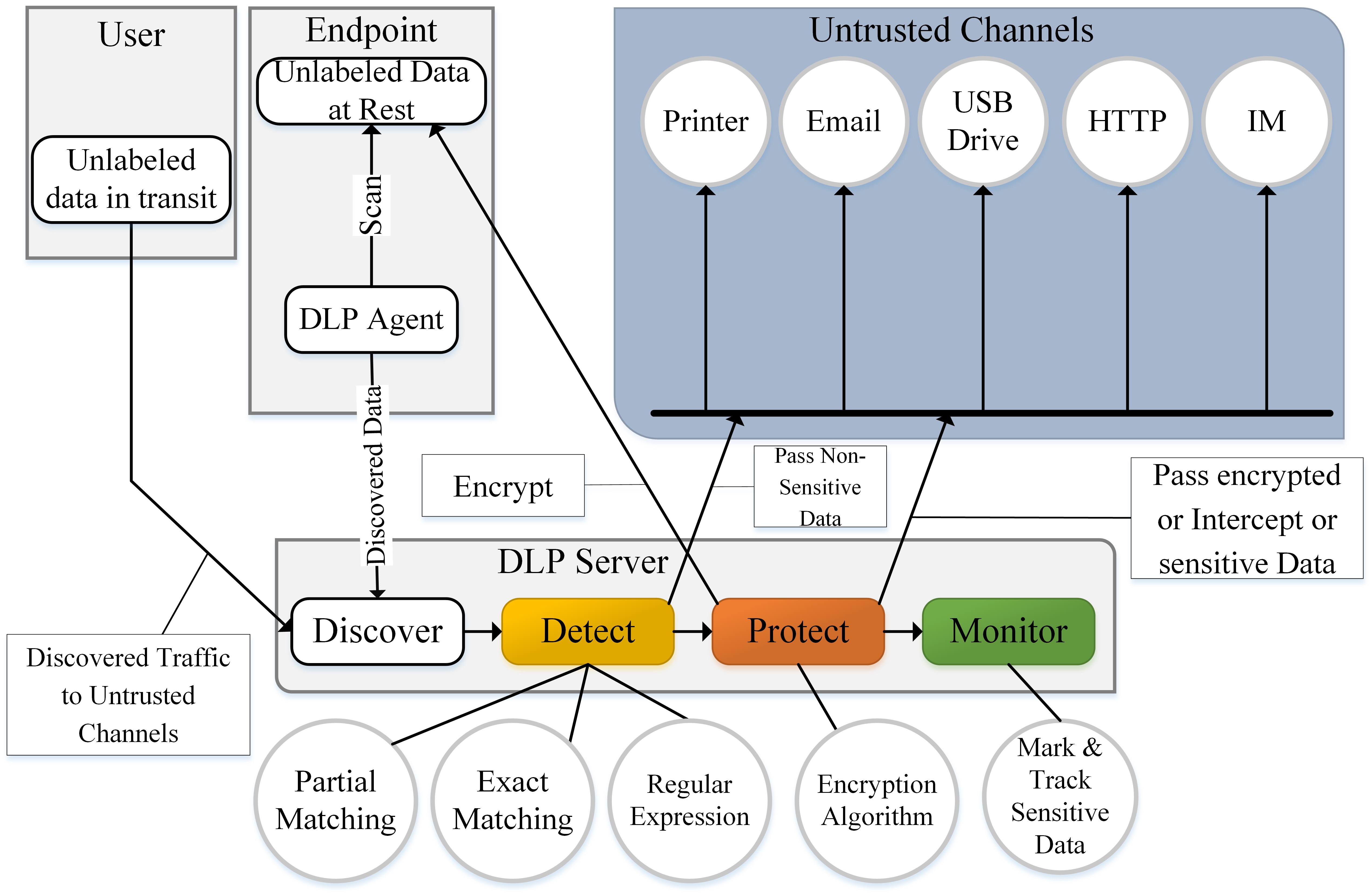}} 
 \caption{\small \textit{ A prototypical Data Leak Prevention (DLP) framework: Data can be \textit{discovered} either through scanning end points for unlabeled data or capturing data in transit to untrusted channels. Data is then inspected by the \textit{detection} module to determine its sensitivity level.  Data that is deemed non-sensitive is normally passed to untrusted channels and non-sensitive data at rest is marked as non-sensitive. Detected sensitive data is passed to a \textit{protection} module. \cite{Alzhrani-isi16}. }}
\label{fig:dlp}
\end{figure}

The aforementioned incidents also have a troubling commonality: namely, they were perpetrated by insiders. 
While data leaks can be perpetrated by both external parties and insiders, insiders are particularly problematic, since they have seemingly legitimate access to sensitive data. 
This renders them immune to preemption by conventional cybersecurity safeguards like firewalls and other access control mechanisms.
Moreover, data leaks by insiders are not necessarily malicious -- they can also occur as a result of human error, so even perfect security vetting cannot prevent all insider-based leaks.

While characteristics of recent data leaks and the threats posed by insiders can be recounted at great length, little else needs to be said to convey the demand for effective data leak prevention (DLP) systems.
Commercial DLP has received some attention lately, e.g., \cite{katzer2013office,ouellet2011magic}, due to the increasing number of leaks, and there are four broad components to an effective commercial DLP system: discovery of what data might be vulnerable to leakage, detecting senesitive data and annotating it, protecting the annotated data, and finally monitoring it (cf. Fig. \ref{fig:dlp}).
To date, much work has been devoted to tracking and securing annotated text, but little work has been devoted to raw detection of sensitive text. That is the direction that we pursue in this paper. 

Particularly, we build prior work in which we were the first to perform evaluation on actual sensitive text data acquired from the Wikileaks dataset.
In these works, we introduced two different techniques -- one based on local learning of classifiers across clusters and a second which, along similar intuition, uses topic models to prune the training set. 
While we compared these results to baseline machine learning algorithms, e.g., SVMs, Naive Bayes, and Logistic Regression, and achieved statistically superior results, the experimental protocols were heterogeneous, and to maintain scope of each paper, we did not provide a comparison of our two novel approaches.  
In this paper, we compare both approaches under one unified evaluation protocol. 
We show that the method of doing local learning on clusters outperforms the dataset pruning method, which in turn outperforms all baseline classifiers.

\section{Related Work}

The problem of sensitive text classification is a sub-problem of natural language processing (NLP) and machine learning.
The simplest methods consist of matching keywords or phrases. 
Additional NLP pre-processing techniques like stemming and stop-word removal can be applied to enhance accuracy. 
More sophisticated NLP rules, like regular expressions and context-free grammars \cite{viola2005learning} can also be applied standalone or in conjunction with keyword/phrase match. 
While keywords offer high recall, they also offer lower precision than such rule-based approaches. 
Rule-based approaches can be used to enhance precision, although rule-based approaches alone tend to result in diminished recall, since matching a keyword is a less exacting constraint than matching NLP rules. 
A shortcoming of pure NLP rule-based approaches like formal grammars and regular expressions, however, is that they are limited in the manner that they generalize to novel data types. 
This is the motivation for employing machine learning algorithms, which, based on training samples, attempt to parameterize a decision boundary with which to make hypotheses about query samples. 
We pursue machine learning approaches in this paper, particularly feature space models, in which a text sample is transformed into a vectorized representation and a hypothesis function, parameterized over the training set, is used to make the classification decision. 
State space models\cite{rudd2016survey}, which analyze state transition based on sequences of input data, and give an output classification similarity/dissimilarity over the sequence of state transitions are another approach, but we do not examine them in this work.
NLP rule-based approaches and machine learning each have their own respective advantages and disadvantages, and a commercially viable solution would likely consist of some combination thereof.

We are not the first to explore applying machine learning toward sensitive text classification.
However, previous authors, e.g., \cite{alneyadi2014semantics,hart2011text,gomez2010data}, have discussed the problem and proposed solutions for approaching it.
Unfortunately, real sensitive text datasets are generally kept private, so these works were not able to perform realistic evaluations, instead using fictitious ``sensitive'' texts (e.g., collected from public Twitter feeds).
To perform a more realistic evaluation of the problem, we introduced the first public sensitive text dataset in \cite{Alzhrani-isi16}, consisting of diplomatic cables made public by the WikiLeaks organization, and performed evaluations using a novel system design -- \textit{Automated Classification Enabled by Security Similarity} (ACESS) -- which performs local classifier learning over clusters of similarity groups.
This work, while seminal, had the disadvantage that the protocol was inherently expensive to tune hyperparameters over and had no separate validation set. 
We later introduced a different approach and a novel partitioning of our Wikileaks dataset in \cite{alzhrani2016automated}. 
This approach uses topic modeling, assesses topic purity, and prunes impure topics from the training set. 
Like ACESS, it achieves state-of-the-art performance on the Wikileaks dataset over baseline classifiers. 
In this paper, we compare ACESS with our pruning approach.

\section{Text Security Classification : Document vs. Paragraph}\label{DocumentVParargraph}
In this section, we define text security classification  and illustrate the difference between paragraph-based and document-based based security classification. 
While sensitivity level could be assessed at even lower granularity, doing so would be analogous, e.g., at the sentence level vs. paragraph level.
The important point is that in reality, for many documents deemed ``sensitive'', only a fraction of their text -- sometimes a tiny fraction -- is actually sensitive.

\subsection{Feature Set}
 Unlike regular text classification, few infrequent features can change the whole document class to a higher security level. Features extracted from the document level differ from features extracted from the paragraph level. Security classification is defined  as follows. Each document $D_i$ has one or  $n$  informative sections or simply paragraphs $P$,   
$D_i = \{P_{1}, \dots, P_{n} \}$. 
 Paragraphs consist of multiple terms or features $F$. In document-based classification,  one feature set is derived for all the paragraphs, this features set determines the document security class   
  $D_i^{Class}  = \{P_{1}, \dots, P_{n} \} =  (F_{1}, \dots, F_{n} )$.

The downside of document-based approach is the assumption that all document portions belong to the same security class which is unlikely. Consequently, the document feature set will accommodate irrelevant features assigned to the document class. Predicting the document class in document-based classification is straightforward compared to paragraph-based. In paragraph-based classification, a  document consists of  paragraphs with one or more different security classes. In order to avoid inconsistency, the security classes described in this section match the classes that appeared in the dataset used in our experiments. The security classes are Unclassified, Confidential and Secret which is  defined respectively  as follows $Class = \{U, C, S\}$. Let ${\cal C}(P)$ be the operator that returns the classification of paragraph $P$.

The class set sorts the security classes from the lowest security level as the first member in the set to the highest. In paragraph-based security classification, a document has multiple feature sets, one set for each paragraph within the document. The security of the paragraph is determined based on its extracted features $ D_i =  {\cal C}(P_i), \dots, {\cal C}( P_{n}) \} =\{ M(f_{1,1}, \dots, f_{1,m} ), \dots, M(f_{n,1}, \dots, f_{n,m})\}$, where $M$ is the machine that estimates the class, $m$ is number of features in each paragraph and  $f_{i,j}$ is feature $j$ from paragraph $i$. This approach reduces the number of irrelevant features correlated to the respective security class. Paragraphs' security classes are determined independently. The document is labeled with the  highest security class of any it's paragraphs ${\cal C}(D_i) = max_j \ {\cal C}(\{P_j)\}$. Thus a document labeled with the  security class $S$  could have paragraphs with any class while a document $D_i$ labeled unclassified must have ${\cal C}(\{P_j)=U, \forall P_j\in D_i$.

\subsection{Classification Probabilities}\label{sssec:num1}
While adapting a powerful classifier with the appropriate features is essential for reducing the error rate, there are other factors that play a critical role in the final outcome such as class representation and number of classes. Paragraph based security classification is a special case where a single paragraph in a document can change the whole document class; Therefore, understanding paragraph class influence over the document class would provide clear insight into the difficulty of paragraph classification process. 

Most researchers address the problem of text classification from the whole document or article point of view. Assuming a uniform prior over classes, i.e., that all classes are balanced without respect to the classifier's performance, the prior probability of a document belonging to one of the classes is $ Pr(1) = 1/2$. However, for three classes the uniform prior probability of one document belonging to one of them is reduced to $Pr(1) = 1/3$ which gives a clear indication that in general multi-class classification problems are harder than the binary single class detection problem. 

As for paragraph-based security classification where the probability of a document belonging to one of the security levels is solely dependent on its paragraph classes. Considering the  $Class$ set in the previous subsection, a document with the label $U$ requires all the paragraphs within the document be labeled as Unclassified, thus the prior probability of a document belonging to class $U$ is  $Pr(D^{U})  =Pr(P_1^{U}) \times \dots \times Pr(P_{n}^{U}) = (1/3)^n$ where $n$ is  the number of paragraphs in the document. It is worth noting that the position of the paragraph does not impact the document class, whereas the probability of a document belonging to the $C$ class is the probability of only one paragraph classified as Confidential regardless of its position and the remaining paragraphs  are classified as either $C$ or $U$.  $Pr(D^{C}) = Pr(P_j^{C}) + (Pr(P_1^{C} | P_1^U) \times \dots \times Pr(P_{n}^{C} | P_{n}^{U})) = 1/3 + ( (2/3) \times \dots \times (2/3))$. assuming a uniform prior. Compared to Confidential and Unclassified, and again -- under a uniform prior, documents have  higher probability to be labeled as $S$ because Secret documents can have paragraphs with any class as long as there is one Secret paragraph within the document $Pr(D^{S}) = P_j^{S} = 1/3$. Hence, the more paragraphs there are in the document, the more feasible it is that the document is labeled as Secret. Also, note that decreasing the number of security classes would improve the classification probability, though the probability calculation presented earlier did not consider two influential factors: namely, the posterior has little to do with class frequency and much to do with the classification algorithm, and that a uniform prior on class frequency does not hold when classes are unbalanced. Nonetheless, considering the case of a uniform prior is important for understanding the intuitive differences between document-level classification considerations and paragraph-level considerations.


Even though paragraph-based classification enables information access to  unclassified or non-sensitive paragraphs  within documents  that contains security information, the documents whole class is totally dependent on the paragraphs' classes. As formerly stated, a document is labeled with the highest security class assigned to its paragraphs. In security text classification, we can build a classifier that is skewed toward the Secret or Confidential class over the Unclassified class. By following this method, the number of misclassified paragraphs as Secret or confidential will also misclassify the document that originally should be labeled  from Unclassified to Confidential or Secret. If information owners prefer to implement an access control mechanism based on document rather than paragraph class, then skewing the classifier to the highest security level will limit access flexibility.

\begin{figure*}[!th]
    \centering
    \subfloat[Impure Topic: This topic is highly mixed with Secret and Confidential paragraphs. ]
{{\includegraphics[width=4cm,height=3cm]{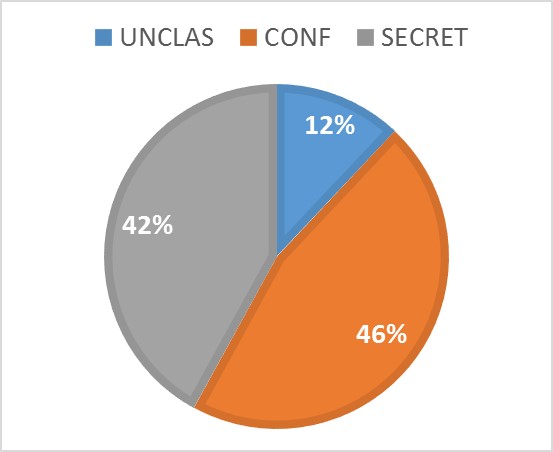} }}%
    \qquad
    \subfloat[Impure Topic: This topic is highly mixed with Unclassified and Confidential paragraphs.]
{{\includegraphics[width=4cm,height=3cm]{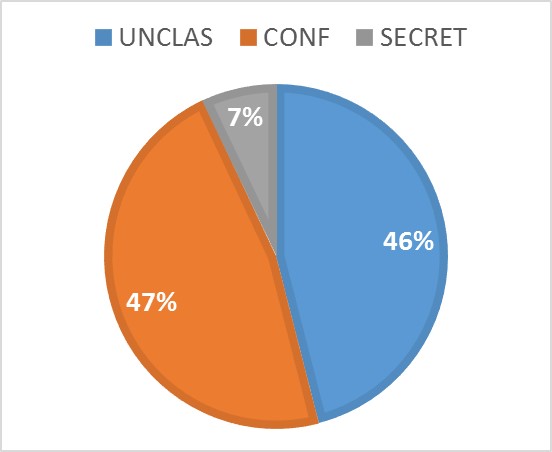} }}%
	\qquad
    \subfloat[Pure Topic: This topic is  populated mostly by Confidential paragraphs.]
{{\includegraphics[width=4cm,height=3cm]{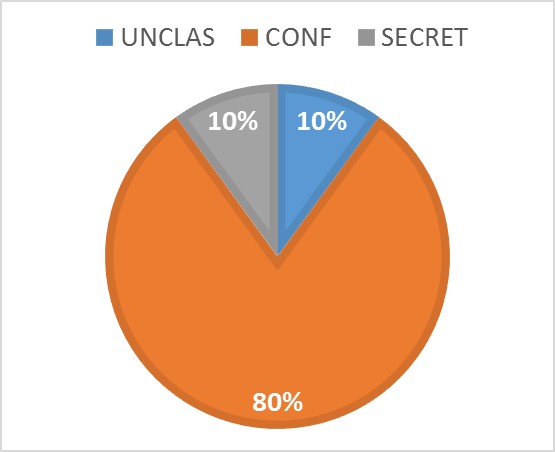} }}%
    \caption{\small \textit{ Examples of topic purity based on Berlin thresholds \cite{alzhrani2016automated}}}%
    \label{fig:TopicPurity}%
\end{figure*}

\section{Insights into WikiLeaks Dataset}

In this section, we provide a description and observations about the U.S diplomatic cables dataset, dubbed the WikiLeaks Dataset. The processes taken to reconstruct the dataset are discussed \cite{Alzhrani-isi16},\cite{alzhrani2016automated}.  
Each U.S diplomatic cable consisted of three sections. The first section is the head section, which includes some information regarding the sender and the receiver of the cable, cable security class, reference number and other information. The creation date of the cable is also included in the head section. We made sure when we created the dataset that we preserve the date, origin of the cable in the paragraph ID. Only the year and month of each cable is added to the paragraph ID. 

In \cite{alzhrani2016automated}, we explained the security classification system utilized in labeling the U.S diplomatic cables. The cable itself has a label in the head section and its information sections or the paragraphs are also labeled. The cable label was in full words such as "SECRET", while the paragraph labels were between brackets in single capital letters such as "S". The document label is one of the entities that paragraph IDs consisted of. In addition to that, we kept the cable number and the paragraph position in the cable in the paragraph ID. 
The second section was the subject, and finally the third one was the body of the cable itself which included one more information sections which we also call a paragraph. 

The embassies: Baghdad, London, Berlin and Damascus were chosen randomly from many other embassies. However, dataset corresponding to the Baghdad embassy had, by far, the largest number of cables. The total number of instances in a single dataset can be up to 50,000 paragraphs. Some cables especially the UNCLASSIFIED ones, are used as templates for daily reports, therefore, many features are shared among them. 

\begin{figure*}[!th]
\centering
{\includegraphics[scale=.6]{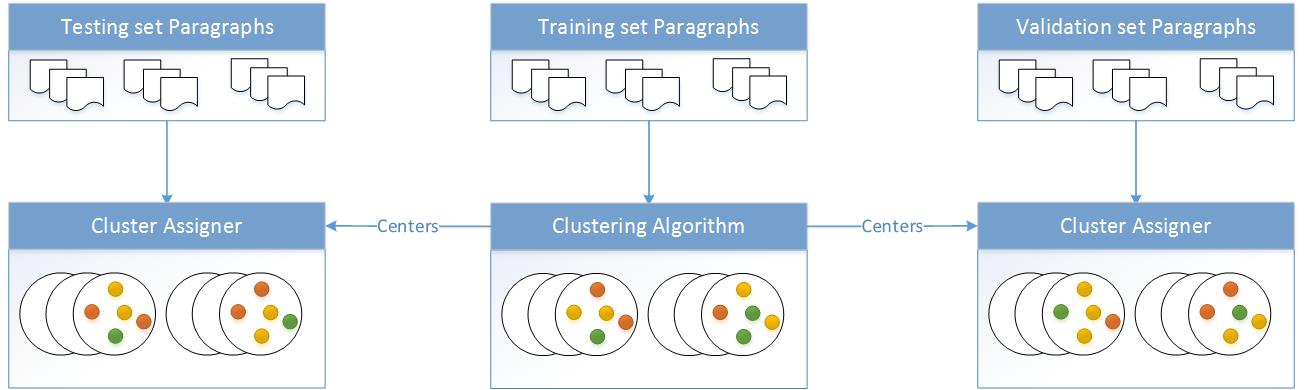}} 
 \caption{\small \textit{ This figure does not show the intermediate pre-processing procedures.  First, similarity clusters are generated out of the training set. Clusters models are utilized to assign both validation and testing set paragraphs to one of the training set clusters. The paragraphs for training, validation and testing clusters are kept in separate files.   }}
\label{fig:cluster}
\end{figure*}

\section{Methodologies}
In this section, we will briefly outline the Topic Model Pruning algorithm and describe in detail the new partitioned version of ACESS.

\subsection{Topic Models Pruning}

The topic model pruning algorithm is described in  \cite{alzhrani2016automated}. In short, the approach locates and removes impure topics by examining the purity of main topics and subtopics generated from the training set. 
The purity threshold lower and upper values are selected based on the classes percentages in the training set. 
In Fig. \ref{fig:TopicPurity}, three graphs illustrate topic purity examples based on the Berlin thresholds. 
The topic is considered impure if the two of the closest class's percentages (SECRET \& CONF or UNCLAS \& CONF) within the topic fall between the predefined thresholds' lower and upper values with respect to the class. 
The security classification levels are hierarchical, which means Secret is more sensitive than Confidential and Confidential is more sensitive than Unclassified.
Since Confidential is between Unclassified and Secret in the sensitivity hierarchy, we would expect this to be the class of highest confusion. 
Therefore, we did not examine Unclassified and Secret class purities -- only Confidential. 
Topics are generated twice, the purpose of the first run is to extract the main topics. 
Instances that belonged to the two conditioned classes are extracted and retain the class's instances which are not conditioned. The second run is to extract the subtopics from all the extracted instances. The same purity conditioning is applied and this time, and the instances are permanently removed.     

\subsection{Automated Security Classification Enabled By Similarity (Partitions Version)}

In the ACESS paper \cite{Alzhrani-isi16}, we generated a wide range of clusters and built a classification model based on each of these clusters. 
The intuition behind this was to find the group of classification models with the best results via a cross-validation technique. 
Looking back at the results from that experiment we found that that, up to a point, the F-Measure results improved as we increased clusters, then they declined.
This experiment itself took a very long time to perform, due to the large number of clusters and classifiers that were built. 
Therefore, our methodology in this paper only generates clusters once for each dataset, and then builds the classification models based on that set of clusters. 
Still, the optimal number of clusters is not known; therefore, we use a number of clusters proportionate to the number of instances in training set.
Also, number of features that we have selected to use this time is different than before. 
In the original version of ACESS, we selected small set of similarity features to make sure that the clusters were populated with paragraphs with different classes. 
In this version, the number instances and class representation does not matter.
A larger set of similarity features is selected and hence fewer paragraphs can populate the clusters. 
This can also be viewed as a cleansing process where an outlier or a single paragraph is assigned to a single cluster by itself. 
Similar to the older version of ACESS, we used a unigram feature to generate clusters.
The following is an illustration of the steps taken in ACESS the partition version:

\subsubsection{Group of Similarity Clusters}

The first step is to generate clusters from the training paragraphs. 
To do so, we first tokenize the paragraphs and extract similarity features from the training set. 
The features with highest TF-IDF values are selected as similarity features. 
The number of features may vary depending on the number of feature values. 
We do not apply any normalization, we remove stop words and do not perform stemming.
The number of clusters that we select corresponds to the number of training paragraphs divided by a default value of 200.
We expect that optimizing this hyperparameter over the validation set can yield better results. 

\begin{figure}[!t]
\centering
{\includegraphics[scale=.4]{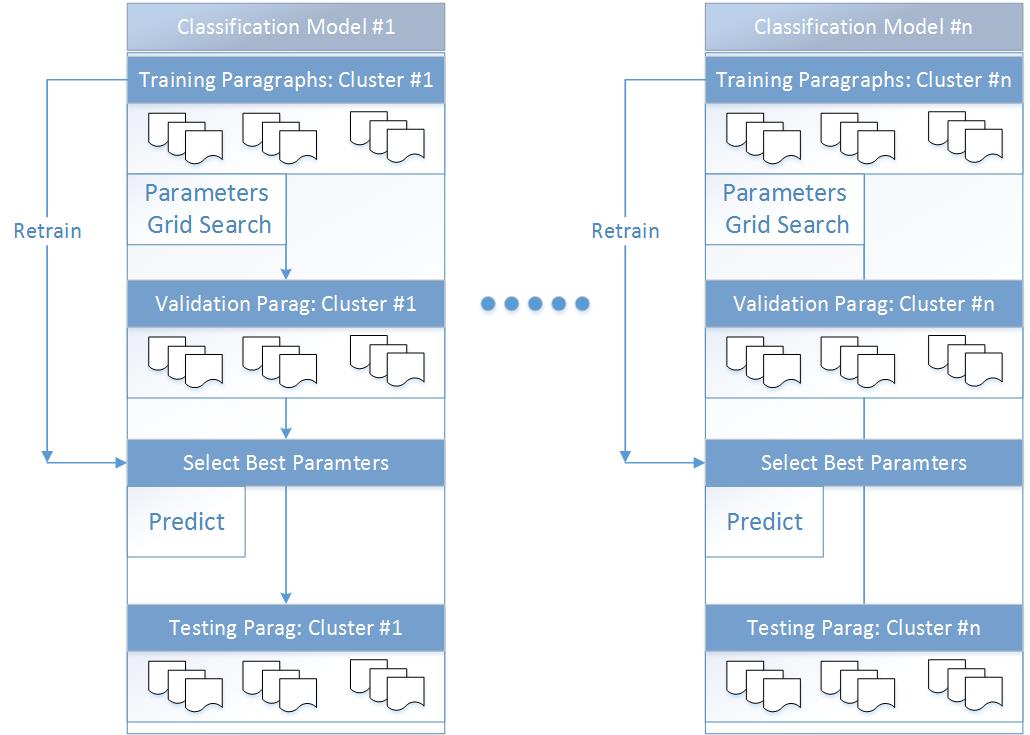}} 
 \caption{\small \textit{The classification models are built over cluster obtained from the training set. 
For each training cluster, optimal classification model hyperparameters are selected according to the points in the validation set that are closest to that corresponding cluster.
The performance of this classification model is evaluated for its corresponding points in the test set.
}}
\label{fig:classification}
\end{figure} 
  
The same pre-processing procedures are reapplied on the validation and testing sets. The clustering model built upon the training set is utilized to assign both validation and testing paragraphs to one of the training clusters. However, each cluster-set is collected separately.
Note that since validation and testing points are assigned to the cluster of nearest training data, some clusters will not be used during validation and testing.

\begin{figure*}[!th]
    \centering
{{\includegraphics[width=18cm,height=6.1cm]{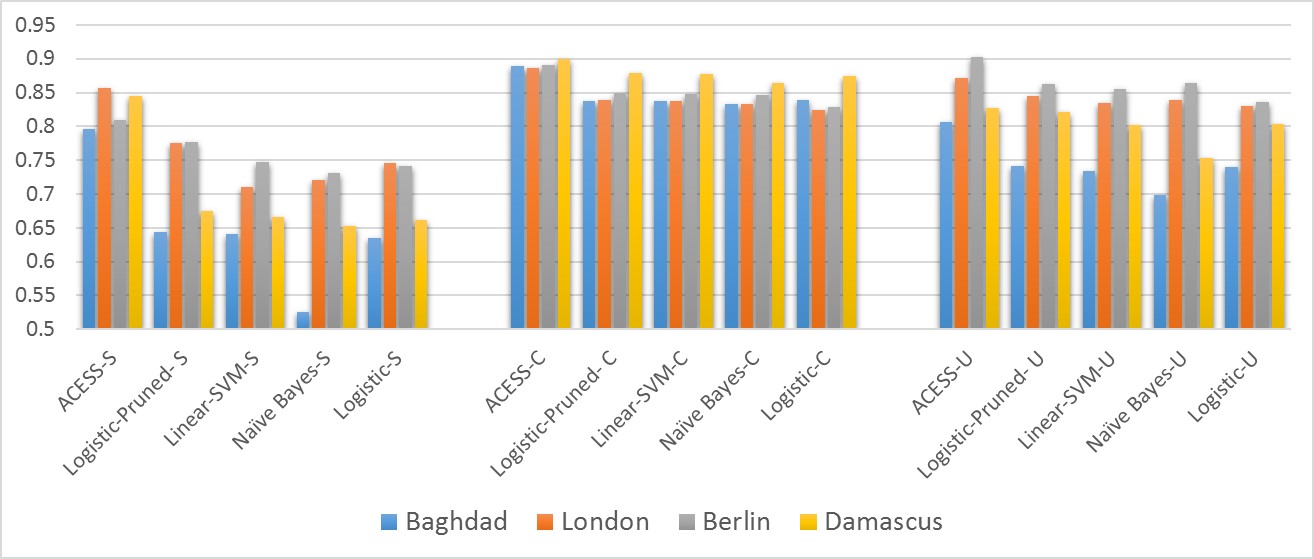} }}%
    \caption{\small \textit{ This figure displays the results obtained from running the baseline algorithms, Logistic Regression with pruning and ACESS. ACESS outperformed all the baseline algorithms and Logistic Regression on the pruned data. The lowest F1-Measure achieved across all the classes is over 0.5, therefore, the y axis starts from 0.5. S, C, and U are initials for Secret, Confidential and Unclassified.  }}%
    \label{fig:ResultsAll}%
\end{figure*}  
 
\subsubsection{Group of Classification Models}
The number of classification models generated for each dataset is the same as the number of training clusters, with one classifier per cluster. 
Classifier parameters are tuned on the validation set. 
The same set of security features is selected and examined against the validation set (cf. Fig~ \ref{fig:classification}). 
Similar to the clustering process, paragraphs are tokenized, but document frequency is used to value the extracted features, Not TF-IDF. 
Only unigram of alphabetic features is considered. 
Finally, the classifier is retrained with best set of parameters and the results obtained from predicting testing paragraphs are reported. 

\section{Experimental Evaluation}

\subsection{Evaluation Metric} 
 Depending on the nature of the organization, the damage resulting from leaking sensitive information might be more significant than interrupting services or restricting eligible users from accessing non-sensitive information. In this case, detecting sensitive paragraphs must be prioritized over service convenience. 
However, this paper assumes that all security classes are equally important, to balance between risk and convenience. 
Therefore, we use F1-score over precision and recall as a performance measure to compare between different prediction models. 
We report F1-Measure per class -- the harmonic mean of precision and recall.
Where $TP$, $FP$, and $FN$ are the number of \textit{true positives}, \textit{false positives}, and \textit{false negatives} respectively, this measure is defined as
\[	
   \textstyle
	\text{F1-Measure} = \frac{2 \times TP}{2 \times TP + FP + FN}.
\]
 
\subsection{Baseline}
 
 The baseline models that we compare to in this paper are Linear-SVMs, Naive Bayes, and Logistic Regression without a pruned dataset. 
We compare our ACESS model and Logistic Regression with pruning, which outperformed our baseline classifiers in previous work, \cite{Alzhrani-isi16}, we also attempted using kernel SVMs, but found that there was little gain over linear, perhaps due to the already sparse feature space.
We selected hyperparameters for these classifiers via a grid-search on the validation set, using a TF-IDF feature space, optimized with respect to the maximum number of features to keep and normalization method. The feature space optimization was also performed via a grid search.
For the pruned data, multi-class classification one-vs-one Logistic Regression with a CG solver was used. The TF-IDF feature space was similarly optimized via a grid search over the validation set. We retrained the classifier using the best feature space and report our results on the test set.  Classes were re-weighted to reflect their actual sizes in the new training set. 
The partitioned version of ACESS used a Linear-SVM with grid search over the validation set to obtain each separate $C$ parameter and document frequency vectorization. 
The top thousand features with with highest frequency were selected.
Note that the real number of selected features is greater than a thousand because the feature with least frequency equals to other features with same frequency. The security features are both unigrams and bigrams. 
 
 \subsection{Clustering Setup}
 
\begin{table}[!h]
\tiny
\caption{Statistics for the clustering step.}
 \label{numclust}
  \resizebox{\columnwidth}{!}{%

\begin{tabular}{|c|c|c|}
\hline
\centering\textbf{Dataset}   & \centering\textbf{ \#  clusters}  & \centering\textbf{ \# features}\cr
\hline Baghdad &  200 & 1360   \\
\hline London &  24  & 1645   \\
\hline Berlin &  38  & 1577  \\
\hline Damascus &  33  & 1549   \\
\hline
\end{tabular}
}
\end{table}

Unlike the experiment in the ACESS paper, one set of clusters is constructed. The table \ref{numclust} displays number of clusters for each dataset. In the similarity clustering step, the top thousand features with highest TF-IDF were selected, note that the real number of features is larger than 1000, because multiple features equal the lowest TF-IDF value. 
The exact number of selected features is in table \ref{numclust}. 
Finally, the initialization for the K-means clustering was random.

\subsection{Results}

Fig. \ref{fig:ResultsAll} displays respective classification results on each respective dataset. We see that the partitioned version of ACESS outperforms all other baseline algorithms as does Logistic Regression on pruned data. 
However, ACESS outperforms our pruned version of Logistic Regression.

\section{Conclusion}

In this paper, we have demonstrated that ACESS outperforms topic model pruning as well as other baselines.
This evaluation demonstrates how well respective classification models work for sensitive text detection over several real-world sensitive text datasets and gives us intuition about what constitutes a good model. 
ACESS and our topic model pruning approaches are fundamentally different; perhaps we can extend this work by fusing concepts from each approach.
Much additional future work can also involve components of the algorithms/data acquisition that are not directly related to the classifier, including gathering additional datasets, e.g., with other leaked Wikileaks data from the same source or others (e.g., Panama papers), and using a better feature space representation, e.g., a deep neural network vectorization algorithm like Word2Vec.

\bibliographystyle{IEEEtran}
\bibliography{IEEEabrv,ref}

\end{document}